\documentclass[conference]{IEEEtran}
\IEEEoverridecommandlockouts
\usepackage{graphicx}
\usepackage{booktabs} 
\usepackage{courier}
\usepackage{graphicx}
\usepackage{stfloats}
\usepackage[all]{xy}
\usepackage{MnSymbol}
\usepackage{amsmath}
\usepackage{multirow}
\usepackage{stackengine}
\usepackage{fmtcount}
\usepackage{amsmath}
\usepackage{array}
\usepackage{algorithm,algorithmicx,algpseudocode}
\usepackage{makecell}
\usepackage{tabularx}
\usepackage{booktabs}

\usepackage{tikz}
\usepackage{array}
\usepackage{booktabs,adjustbox}

\makeatletter
\def\endthebibliography{%
	\def\@noitemerr{\@latex@warning{Empty `thebibliography' environment}}%
	\endlist
}

\makeatother

\def\BibTeX{{\rm B\kern-.05em{\sc i\kern-.025em b}\kern-.08em
    T\kern-.1667em\lower.7ex\hbox{E}\kern-.125emX}}
\begin{document}

\title{Finding Cryptocurrency Attack Indicators Using Temporal Logic and Darkweb Data}
\author{
	\IEEEauthorblockN{Mohammed Almukaynizi$^{1}$, Vivin Paliath$^{2}$, Malay Shah$^{2}$, Malav Shah$^{2}$, Paulo Shakarian$^{1,2}$}
	\IEEEauthorblockA{$^{1}$Arizona State University \\
		$^{2}$Cyber Reconnaissance, Inc. \\
	Tempe, AZ, USA \\
	\{malmukay, vivin, msshah11, mpshah5, shak\}@asu.edu}
}

\maketitle
\thispagestyle{plain}
\pagestyle{plain} 

\IEEEpubidadjcol

\begin{abstract}

With the recent prevalence of darkweb/deepweb (D2web) sites specializing in the trade of exploit kits and malware, malicious actors have easy-access to a wide-range of tools that can empower their offensive capability.
In this study, we apply concepts from causal reasoning, itemset mining, and logic programming on historical cryptocurrency-related cyber incidents with intelligence collected from over 400 D2web hacker forums. Our goal was to find indicators of cyber threats targeting cryptocurrency traders and exchange platforms from hacker activity. Our approach found interesting activities that, when observed together in the D2web, subsequent cryptocurrency-related incidents are at least twice as likely to occur than they would if no activity was observed. We also present an algorithmic extension to a previously-introduced algorithm called APT-Extract that allows to model new semantic structures that are specific to our application.

\end{abstract}


\section{Introduction and Dataset}

Cryptocurrencies are digital currencies that mostly use the blockchain concept to record transactions. Perhaps the most well-known one is Bitcoin. 
It was estimated that the market capitalization of cryptocurrencies has exceeded 400 billion dollars, after peaking at over 700 billion dollars\footnote{https://cointelegraph.com/news/combined-crypto-market-capitalization-races-past-800-bln.}.

With the high reliance on technology, increasing adoption from businesses and traders, and due to the inherent anonymity associated with transactions and wallet owners, malicious threat actors (including hackers and scammers) aiming for financial gain have been highly motivated to hack and scam to gain control over cryptocurrency wallets and perform transactions. In this research effort, we have encoded and recorded over 50 major incidents of cyberattacks and fraud campaigns targeting multiple cryptocurrencies and traders\footnote{These incidents were extracted from various IT security repositories including https://www.hackmageddon.com, https://www.cryptoaware.org/resources/notable-cryptocurrency-hacks, https://www.livecoinwatch.com, and others.}. A few examples are reported in Table~\ref{1}. We also queried postings from over 400 hacker forums in the D2web using an API that is commercially provided by a threat intelligence firm\footnote{Cyber Reconnaissance, Inc. (CYR3CON), https://www.cyr3con.ai.}. The database is collected from D2web sites that were identified to be serving hacking-related content. 
Using both databases, we mine for sequential patterns to identify indicators of attacks from the D2web database, which can then be monitored to reason about risks that are likely to occur in the future. Specifically, we found that when certain D2web activities are observed together, subsequent cryptocurrency-related incidents are at least twice more likely to occur than they would if no activity was observed and significantly more likely to occur than when individual activity is observed.
 
\vspace{-1em}
\begin{table}[!h]
	\caption{Example attacks}
	\vspace{-3em}
	\label{1}
	\begin{center}
		\begin{tabular}{|l|c|c|}
			\hline
			\bf{Name of Attack} & \bf{Targeted Currency} & \bf{Estimated Loss (in USD)}\\
			\hline
			{Mt Gox Hack } & {Bitcoin } & {450 Millions }\\
			\hline
			{NiceHash Hack } & {Bitcoin } & {62 Millions }\\
			\hline
			{Parity Wallet Hacks } & {Ethereum } & {160 Millions }\\
			\hline
			{Tether Token Hack } & {Tether } & {30 Millions }\\
			\hline
		\end{tabular}
			\vspace{-1em}
	\end{center}
\end{table}

To learn such patterns, we sought to derive temporal logic rules of the form ``A cryptocurrency \textit{G} will be targeted by hackers/scammers with a probability \textit{p} within $\Delta t_{act}$ time units after itemset \textit{F} is observed on the D2web within $\Delta t_{con}$ time units". To do so, we combine concepts from logic programming (in particular, the concepts of existential frequency function (\textit{efr}) from APT logic~\cite{Shakarian:2011, Shakarian:2012}) with frequent itemset mining~\cite{han2000mining} and temporal causal reasoning~\cite{kleinberg2009temporal}. Table~\ref{2} shows some of the interesting \textit{efr} rules obtained from the algorithm.

\begin{table*}
	\centering
	\caption{Example of significant rules}
	\label{2}
	\renewcommand{\arraystretch}{1.5}
	\begin{tabular}{|m{6cm}|c|m{6cm}|}
		\hline 
		\textbf{Rule} & \textbf{\%increase in likelihood of occurrence} & \textbf{Description} 
		\\ \hline 
		\textit{Mentioned(Coinbase, financial, 10)} $\wedge $    \textit{Mentioned(Gmail, software, 13)}  $ \overset{efr}{\rightlsquigarrow}$
		~~~~~~~~~~~\textit{Attacked(general)}: [21, 0.94, 5]
		& 201\% & When ``Coinbase" as a financial tag and ``Gmail" as a software tag are mentioned more than 10 and 13 times, respectively, within any window of 5 days, there is a 94\% chance that a cryptocurrency-related incident would happen in the next $21$ days. This is 201\% more likely to occur than the chance of occurring in any sequence of 21 days. 
		\\
		\hline  
		\textit{Mentioned(SQL, software, 19)} $ \wedge $  ~~~~~~~~~~~~~~~\ ~~~~~~~~~~~\textit{Mentioned(Windows Registry, software, 19)} $\overset{efr}{\rightlsquigarrow} $ \textit{Attacked(Ethereum)}: [21, 0.39, 5] 
		&206\% &  When ``SQL" as a software tag and ``Windows Registry" as a software tag are mentioned more then 19 times, there is a 39\% chance that a Ethereum event would happen in the next 21 days, 206\% increase in likelihood of occurrence. \\ 
		\hline 
	\end{tabular} 
\end{table*}

	\vspace{-0.25em}
\section{Technical Approach}

\vspace{-0.5 em}
\subsection{APT-Logic} 
\noindent\textbf{Concepts.} We use the same syntax and semantics that were previously introduced in our work~\cite{Shakarian:2011, Shakarian:2012, stanton2015mining}. Here, we informally review a subset of those concepts. 

A \textit{thread} is a sequence of \textit{worlds} (events). Each world corresponds to a discrete time point. Time points are represented by natural numbers in the range 1, \dots , $t_{max}$. A \textit{formula} $F$ can be any itemset (in this paper, atoms can only be predicates with arities of 2 or 3, e.g., \{\textit{attacked(Bitcoin)},\textit{ mentioned(Coinbase, software, 9)}, \textit{mentioned(Gmail, software, 13)}\}). The atoms used in this study are partitioned into two disjoint subsets: condition atoms from D2web activities ($\mathcal{A}_{con}$), and action atoms from the database of observed incidents ($\mathcal{A}_{act}$).\smallskip

\noindent\textbf{Satisfaction.} We say a thread ($Th$) satisfies a formula $F$ at a time-point $t$ (denoted $Th(t) \models F$) if and only if:  $\forall_{a \in F} (Th(t) \models a)$. \smallskip

\noindent\textbf{Rule probability.} The probability is determined based on the fraction of times where a rule $r$ is satisfied from the times where the precondition is satisfied\footnote{If the precondition is not satisfied by \textit{Th} at any time-point, \textit{p} to set to 1.}. \smallskip

\noindent\textbf{Significance.} We determine whether a rule is statistically significant based on relative likelihood, i.e., the percentage increase in likelihood of occurrence when the precondition is observed.\smallskip

\subsection {D2web Activity} 
\noindent\textbf{Tagging.} The D2web API supplies tags with each post. Each tag should belong to one of three categories: financial, software, or general topic tags. The tagging algorithm leverages document similarity techniques on doc2vec representations of posts to assign tags that are most relevant to each post. Our approach uses the count of tags per day, i.e., when a tag is observed with a count that exceeds certain threshold, an atom respective to that condition becomes true. The said threshold is tag-dependent; meaning each tag has a threshold. We determined thresholds based on average and standard deviation of tag mentions per day in the historical data\footnote{An atom becomes true if, on a given day, its respective tag is mentioned more than the average added to (1.5 * standard deviation).}.\smallskip



\noindent\textbf{Sliding Window.} The current semantic structure of APT-logic does not capture the concept of \textit{efr} whose precondition atoms occur in any order within a sequence of $\Delta t_{con}$ time points. However, the \textit{efr} rules we sought to obtain need such semantics. Therefore, we make a new thread $Th'$ that is derived from $Th$ according to Algorithm~\ref{alg1}. Essentially, this algorithm iterates over $Th$ once; at each time point $t$ it assigns $Th'(t)$ the condition atoms that $Th(t_i)$ satisfies for every $t_i$ in the sequence of time-points $t- (\Delta t_{con} - 1)$, \dots, $t$.\smallskip

\begin{algorithm}
	\caption{Forward rolling condition atoms}
	\label{alg1}
	\begin{algorithmic}[1]
		\renewcommand{\algorithmicrequire}{\textbf{Input:}}
		\renewcommand{\algorithmicensure}{\textbf{Output:}}
		\Require Thread $Th$, an empty thread $Th'$, length of threads $t_{max}$, condition atoms $\mathcal{A}_{con}$, and length of rolling window $\Delta t_{con}$
		\Ensure  Forward-rolled thread $Th'$
		\State $window\_size \gets \Delta t_{con} -1$ 
		\For {$t \in \{1, \dots , t_{max}\}$}
		\For {$t_i \in \{max(1, t-window\_size), \dots , t\}$}
		\For {$g \in \{a \in \mathcal{A}_{con} \ where \ Th(t_i) \models a\}$}
		\State $Assign(Th'(t), g)$  \Comment{assigns $g$ to $Th'(t)$ if $Th'(t) \nvDash g$ }
		\EndFor
		\EndFor
		
		\EndFor\\
		\Return $Th'$ 
	\end{algorithmic} 
\end{algorithm}

\noindent\textbf{Itemsets.} The preconditions of our \textit{efr} rules are frequent itemsets, obtained from running the Apriori algorithm~\cite{han2000mining}. The input to the Apriori algorithm is a database of transactions\textemdash each world corresponding to a time-point in $Th'$ can be regarded as a transaction. The output is all itemsets that are satisfied by the thread with a frequency exceeding a minimum support\footnote{$min\_support$ is experimentally determined and set to ($0.13 * t_{max}$).}. We run the Apriori algorithm on $Th'$. The algorithm returns frequent itemsets (denoted $freqItemsets$). Then we make a thread $Th_{itemsets}$ only containing atoms (frequent itemsets) that $Th'$ satisfies at each time point as illustrated in Algorithm~\ref{alg2}.\smallskip

\begin{algorithm}
	\caption{Construct $Th_{itemsets}$ }
	\label{alg2}
	\begin{algorithmic}[1]
		\renewcommand{\algorithmicrequire}{\textbf{Input:}}
		\renewcommand{\algorithmicensure}{\textbf{Output:}}
		\Require Thread $Th$, thread $Th'$, an empty thread $Th_{itemsets}$, length of thread $t_{max}$, $freqItemsets$, and action atoms $\mathcal{A}_{act}$
		\Ensure  $Th_{itemsets}$
		\For {$t \in \{1, \dots , t_{max}\}$}
		\For {$I \in freqItemsets$}
		
		\If {$I \subseteq Th'(t)$}
		\State $Assign(Th_{itemsets}(t), I)$ 
		\EndIf	
		\EndFor
		\For {$g \in \{a \in \mathcal{A}_{act} \ where \ Th(t) \models a\}$}
		\State $Assign(Th_{itemsets}(t), g)$
		\EndFor
		\EndFor\\
		\Return $Th_{itemsets}$
	\end{algorithmic} 
\end{algorithm}

%
%
%

\noindent\textbf{\textit{efr} Rule-Learning.} We use the algorithm \textit{APT-EXTRACT} to mine for \textit{efr} rules~\cite{Shakarian:2011}. Essentially, \textit{APT-EXTRACT}  produces all rules whose precondition is any of the itemsets that are satisfied by the thread at $n$ or more time points, i.e., $n$ is greater than a minimum (lower bound) support (specified by the argument \textit{SuppLB}~\cite{Shakarian:2011}).\smallskip

\section{Experiments and Analysis}

We show in this section: (1) the rules that are preconditioned on itemsets tend to have higher correlations with future incidents than the rules that are preconditioned on single atoms, and (2) the performance of our approach change linearly when $\Delta t_{act}$ is extended.\smallskip

\begin{figure}[!h]
	\centering
	\includegraphics[scale=0.15,keepaspectratio]{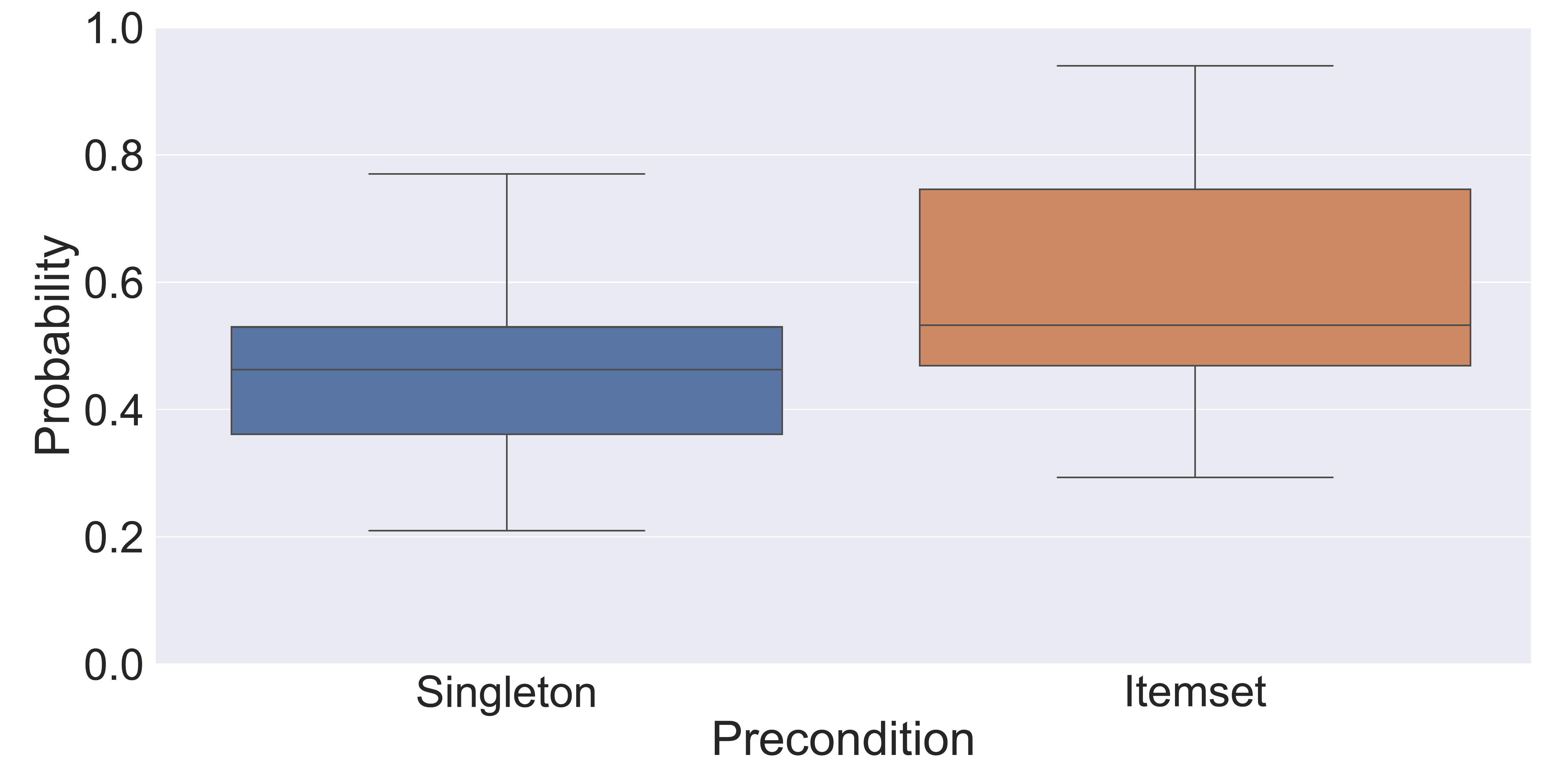}
	\caption{Box plot showing precondition type (itemset rules, single-atom-rules) vs. probability.}
	\label{prob}
\end{figure}

\noindent\textbf{Itemset Rules vs. Atomic Rules.} Interestingly, the \textit{efr} rules whose precondition is itemsets have notably higher probabilities than those whose precondition is single tags. It is evident (from Figure~\ref{prob}) that the usage of the Apriori algorithm  with APT-logic added considerable value to the generated rules. 

\vspace{-0.5em}

\begin{figure}[!h]
	\centering
	\includegraphics[scale=0.15,keepaspectratio]{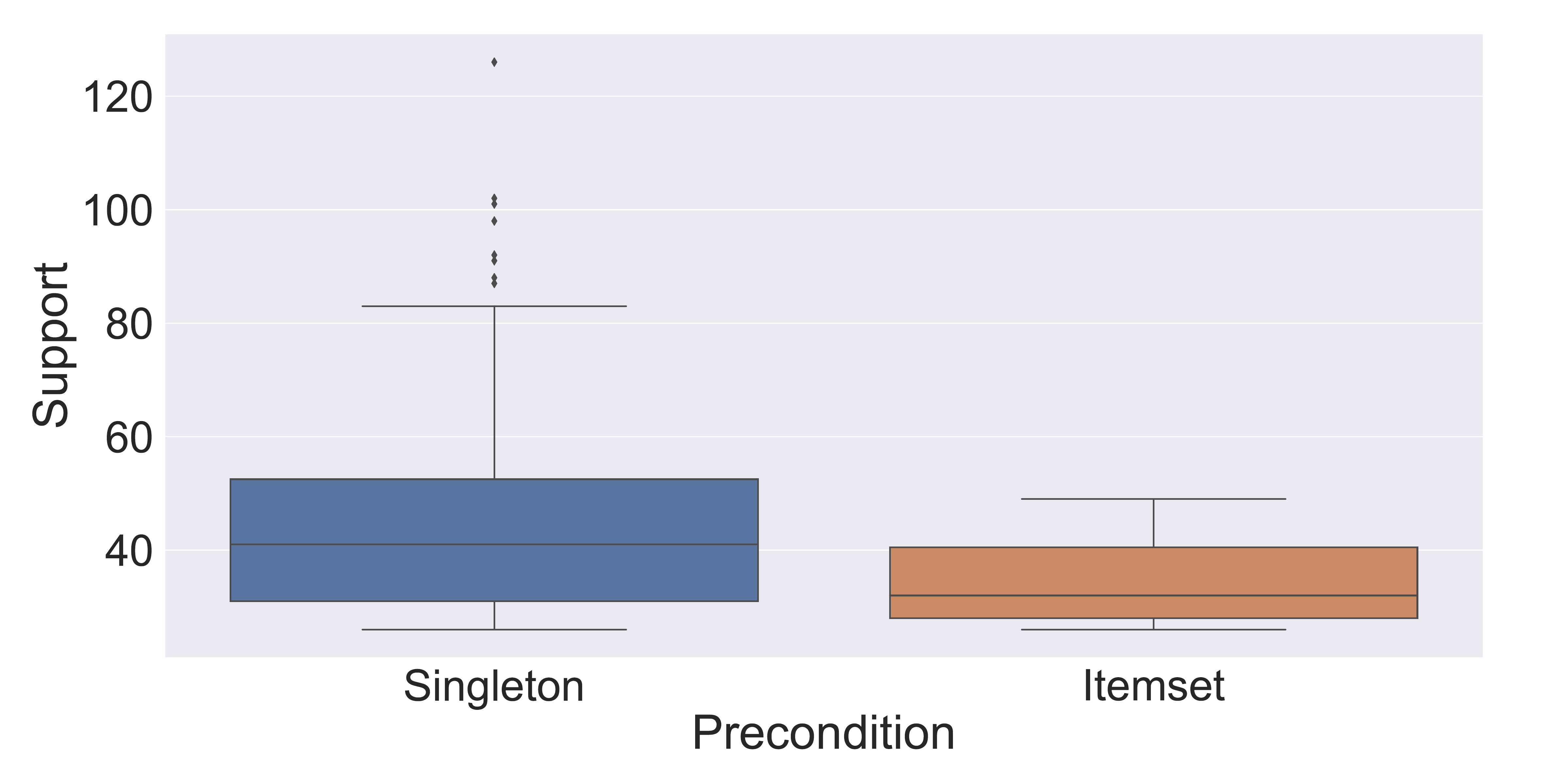}
	\caption{Box plot showing precondition type (itemset rules, single-atom rules) vs. support}
	\label{support}
\end{figure}

Although the single atoms that are used as preconditions of the generated rules may not be in any of the frequent itemsets, the single-atom-rules have on average higher support than the frequent-itemset-rules as depicted in Figure~\ref{support}. This suggests that an atom of an itemset does not necessarily occur when another atom in that same itemset is observed i.e., less co-occurrence. 

\vspace{-1em}

\begin{figure}[!h]
	\centering
	\includegraphics[scale=1.5,keepaspectratio]{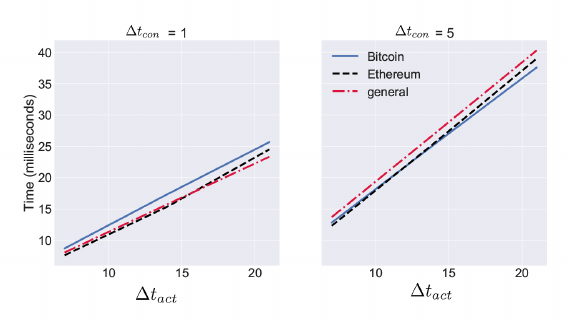}
	\vspace{-1em}
	\caption{Time taken to generate rules when varying $\Delta t_{act}$.}
	\label{fig3}
\end{figure}

\noindent\textbf{Run-time Analysis.} Figure \ref{fig3} shows that when the length $\Delta t_{act}$ increases, the time taken for \textit{APT-EXTRACT} to generate rules increases linearly, because the number of time points for which \textit{APT-EXTRACT} needs to check the satisfaction of the consequence increases. This shows that our approach is flexible to applications where the user wants to extend the length of time-units for which \textit{efr} rules are computed. However, we shall note that the Apriori algorithm runs in exponential time, and for all analysis in this study, we restrict the number of items to 2 in each itemset, and we do not introduce more types of atoms (tags).\smallskip

	\vspace{-1em}

	\vspace{-1em}
\section{Conclusion}

This study presents an approach that combines concepts from causal reasoning, itemset mining, and logic programming. Our approach identifies indicators of cyber threats to cryptocurrency traders and exchange platforms from hacker activity in D2web. Examples of interesting rules are presented in this paper. In the future, we plan to use the approach with other sources such as social media platforms. Additionally, we look to incorporate the learned rules into an alert system that generates and visualizes warnings.

\section*{Acknowledgment}
Some of the authors were supported by the Office of Naval Research (ONR) Neptune program.  Paulo Shakarian, Vivin Paliath, Malay Shah, and Malav Shah are supported by the Office of the Director of National Intelligence (ODNI) and the Intelligence Advanced Research Projects Activity (IARPA) via the Air Force Research Laboratory (AFRL) contract number FA8750-16-C-0112. The U.S. Government is authorized to reproduce and distribute reprints for Governmental purposes notwithstanding any copyright annotation thereon. Disclaimer: The views and conclusions contained herein are those of the authors and should not be interpreted as necessarily representing the official policies or endorsements, either expressed or implied, of ODNI, IARPA, AFRL, or the U.S. Government.

\bibliographystyle{IEEEtran} 
\bibliography{sigproc}

\end{document}